# Improved End-to-End Dysarthric Speech Recognition via Meta-learning Based Model Re-initialization

*Disong Wang[1], Jianwei Yu[1], Xixin Wu[2], Lifa Sun[3], Xunying Liu[1], Helen Meng[1]*

[1]Human-Computer Communications Laboratory
Department of System Engineering and Engineering Management
The Chinese University of Hong Kong, Hong Kong SAR, China
[2]Department of Engineering, University of Cambridge, UK
[3]SpeechX Limited, Shenzhen, China
{dswang, jwyu, wuxx, lfsun, xyliu, hmmeng}@se.cuhk.edu.hk

## Abstract

Dysarthric speech recognition is a challenging task as dysarthric data is limited and its acoustics deviate significantly from normal speech. Model-based speaker adaptation is a promising method by using the limited dysarthric speech to fine-tune a base model that has been pre-trained from large amounts of normal speech to obtain speaker-dependent models. However, statistic distribution mismatches between the normal and dysarthric speech data limit the adaptation performance of the base model. To address this problem, we propose to re-initialize the base model via meta-learning to obtain a better model initialization. Specifically, we focus on end-to-end models and extend the model-agnostic meta learning (MAML) and Reptile algorithms to meta update the base model by repeatedly simulating adaptation to different dysarthric speakers. As a result, the re-initialized model acquires dysarthric speech knowledge and learns how to perform fast adaptation to unseen dysarthric speakers with improved performance. Experimental results on UASpeech dataset show that the best model with proposed methods achieves 54.2% and 7.6% relative word error rate reduction compared with the base model without finetuning and the model directly fine-tuned from the base model, respectively, and it is comparable with the state-of-the-art hybrid DNN-HMM model.

## 1. Introduction

Dysarthria denotes a set of speech disorders related with neurological conditions and diseases such as traumatic brain injury, stroke, Parkinson's disease or amyotrophic lateral sclerosis, which cause disturbances in muscular control over the speech production [1]. As a result, dysarthric speech exhibits very different acoustic characteristics from normal speech, and automatic speech recognition (ASR) systems trained on normal speech suffer from marked performance degradation when applied to dysarthric speech. The problem of modelling is exacerbated by possibly large inter- and intra-speaker variations inherent in dysarthric speech, and scarcity of dysarthric training data further increases the challenges in the development of robust ASR systems for dysarthric speakers.

To address these problems, recent research efforts [2] have explored the use of end-to-end ASR models pre-trained on large amounts of normal speech data and further adapted into speaker-dependent models which produced encouraging results (see Figure 1 (a)). The key to improve ASR performance lies in the success of transferring the knowledge learned from the normal speech, e.g., phonetic variations [3], to dysarthric speech recognition task. However, dysarthric speech differs from the normal speech in terms of various acoustic characteristics, e.g., unstable prosody, incorrect pronunciations and different phonetic variations. Therefore, mismatches of the statistic distributions in acoustic features between normal and dysarthric speech inevitably limit the generalization ability of the base model.

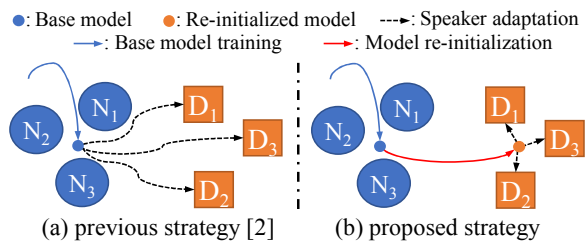

Figure 1. *Illustration of the previous and proposed strategies for model training and speaker adaptation. $N_i$ denotes normal speaker, $D_i$ denotes dysarthric speaker.*

To alleviate this issue, a new strategy is proposed as shown in Figure 1 (b). The idea is to re-initialize parameters of the base model, so that improved speaker adaptation performance can be achieved with the re-initialized model. Inspired by the effectiveness of meta-learning methods applied to various tasks with limited data [4-8], we propose to perform the model re-initialization via meta-learning. Specifically, two meta-learning algorithms, i.e., MAML [7] and Reptile [8], are separately introduced to meta-update the base model by repeatedly simulating adaptation to different dysarthric speakers, which forces the model to learn how to adapt to different dysarthric speakers effectively. The explicit integration of speaker adaptation into the re-initialization process steers the base model to be optimized for generalization. As a result, the re-initialized model acquires speech knowledge across dysarthric speakers, so that the re-initialized model can be fast adapted to unseen dysarthric speakers with improved performance.

The main contributions of this paper are: (1) We propose to extend MAML and Reptile algorithms for model re-initialization and experimentally validate its effectiveness on UASpeech dataset [9] to improve the ASR performance; (2) Per-task batch normalization running statistics are proposed to stabilize the meta-learning process; (3) The proposed MAML and Reptile based model re-initialization methods can be applied to ASR systems with any kind of network architectures.

## 2. Related work

### 2.1. Dysarthric speech recognition

Model adaptation techniques have been successfully applied in dysarthric speech recognition [10]. The maximum a-posteriori (MAP) adaptation in traditional GMM-HMM enables intra-speaker variability modelling to improve ASR performance [11] and enhance the speaker adaptive training (SAT) [12]. Various training and adaptation techniques have also been proved effective in hybrid DNN-HMM systems [13]. Besides, end-to-end architectures have been introduced [2, 14] and reported to significantly improve the accuracy of ASR by adapting the pre-trained base model [2]. In this paper, we focus on end-to-end models to validate the effectiveness of our proposed methods.

### 2.2. Meta-learning

Meta-learning aims to teach a model how to learn various tasks, so that it can solve new tasks with few samples. Hence, meta-learning is suitable for dysarthric speech recognition, especially due to data scarcity in dysarthric speech for speaker adaptation. Recent work has proposed different meta-learning methods [4-8], among which we employ MAML and Reptile for model re-initialization due to their simplicity and effectiveness [7, 8]. Meta-learning has already been used in speech-related fields, such as speaker recognition [15], low-resource and code-switched ASR [16, 17], speaker adaptation for normal and accented speech recognition [18-20]. There is still much room for investigating speaker adaptation for dysarthric speech recognition. However, meta-training an end-to-end model from scratch requires large amounts of data [18-20], and is generally not achievable for dysarthric speech data which is difficult to obtain. Therefore, this paper proposes to meta-train a base model and further explored the use of Reptile as a novelty.

## 3. Proposed methods

In this section, the dysarthric speaker adaptation problem is first formulated. Then two meta-learning methods, i.e., MAML and Reptile, are proposed for re-initialization of a pre-trained base model to obtain a good initialization with improved adaptation performance to unseen dysarthric speakers. In addition, we propose an effective way by using per-task batch normalization running statistics to stabilize meta-learning.

### 3.1. Problem formulation

Assume the number of dysarthric speakers is $I$, and the paired data for $i^{th}$ speaker consists of the adaptation data $(\mathbf{X}_{i,a}, \mathbf{Y}_{i,a})$ and test data $(\mathbf{X}_{i,t}, \mathbf{Y}_{i,t})$, where $\mathbf{X}$ is the speech feature, and $\mathbf{Y}$ is the corresponding character label. Given an end-to-end base model $f(\mathbf{X};\boldsymbol{\theta})$ with the parameters $\boldsymbol{\theta}$ pre-trained with large amounts of normal speech [2], the model adaptation for $i^{th}$ speaker aims to fine-tune $\boldsymbol{\theta}$ over the adaptation data $(\mathbf{X}_{i,a}, \mathbf{Y}_{i,a})$

$$adapt(f, \boldsymbol{\theta}, (\mathbf{X}_{i,a}, \mathbf{Y}_{i,a})) \rightarrow \boldsymbol{\theta}_i \qquad (1)$$

so that the adapted model $\boldsymbol{\theta}_i$ can achieve better performance on the test data $(\mathbf{X}_{i,t}, \mathbf{Y}_{i,t})$. The performance can be measured by the loss function $L(\mathbf{Y}_{i,t}, f(\mathbf{X}_{i,t}; \boldsymbol{\theta}_i))$, which is the character Connectionist Temporal Classification (CTC) loss or Cross-Entropy (CE) loss in the context of end-to-end ASR. Due to the different acoustic characteristics between the normal speech and dysarthric speech, the performance of adaptation to dysarthric speakers by using the base model is limited. To alleviate this issue, we propose to employ MAML and Reptile to re-initialize the base model $\boldsymbol{\theta}$ to obtain a better initialization $\boldsymbol{\theta}^*$ with powerful generalization capacity to unseen dysarthric speakers. The detailed MAML/Reptile based model re-initialization is outlined in Algorithm 1.

---

**Algorithm 1** MAML/Reptile based model re-initialization
**Input:** outer-loop and inner loop optimization steps $K$ and $J$, number of dysarthric speakers $I$, base model $\boldsymbol{\theta}$, adaptation data $(\mathbf{X}_{i,a}, \mathbf{Y}_{i,a})$, learning rate $\alpha$ and $\eta$, meta-learning type

1: $\boldsymbol{\theta}^* \leftarrow \boldsymbol{\theta}$
2: **for** $k \in \{1, 2, ..., K\}$ **do**
3:    **for** $i \in \{1, 2, ..., I\}$ **do**
4:       $\boldsymbol{\theta}_i \leftarrow \boldsymbol{\theta}^*$
5:       Sample training data $(\mathbf{X}^t_{i,a}, \mathbf{Y}^t_{i,a})$ from $(\mathbf{X}_{i,a}, \mathbf{Y}_{i,a})$
6:       **for** $j \in \{1, 2, ..., J\}$ **do**
7:          Speaker adaptation via the gradient descent:
$$\boldsymbol{\theta}_i \leftarrow \boldsymbol{\theta}_i - \alpha \nabla_{\boldsymbol{\theta}_i} L(\mathbf{Y}^t_{i,a}, f(\mathbf{X}^t_{i,a}; \boldsymbol{\theta}_i))$$
8:       **end for**
9:    **end for**
10:    **if** meta-learning type is MAML **then**
11:       Sample validation data $(\mathbf{X}^v_{i,a}, \mathbf{Y}^v_{i,a})$ ($i=1,2,...I$) from $(\mathbf{X}_{i,a}, \mathbf{Y}_{i,a})$ and meta-update $\boldsymbol{\theta}^*$:
$$\boldsymbol{\theta}^* \leftarrow \boldsymbol{\theta}^* - \eta \frac{1}{I} \sum_{i=1}^{I} \nabla_{\boldsymbol{\theta}_i} L(\mathbf{Y}^v_{i,a}, f(\mathbf{X}^v_{i,a}; \boldsymbol{\theta}_i))$$
12:    **else if** meta-learning type is Reptile **then**
$$\boldsymbol{\theta}^* \leftarrow \boldsymbol{\theta}^* - \eta \frac{1}{I} \sum_{i=1}^{I} (\boldsymbol{\theta}^* - \boldsymbol{\theta}_i)$$
13:    **end if**
14: **end for**
15: **return** re-initialized model $\boldsymbol{\theta}^*$

---

### 3.2. MAML based model re-initialization

To apply MAML, adaptation to each dysarthric speaker is treated as a task. First, let the meta-model $\boldsymbol{\theta}^*$ to be initialized with the base model $\boldsymbol{\theta}$. For $i^{th}$ task, i.e., adaptation to $i^{th}$ speaker, training data $(\mathbf{X}^t_{i,a}, \mathbf{Y}^t_{i,a})$ and validation data $(\mathbf{X}^v_{i,a}, \mathbf{Y}^v_{i,a})$ are sampled from the adaptation data $(\mathbf{X}_{i,a}, \mathbf{Y}_{i,a})$. Then the inner-loop and outer-loop optimization are respectively performed for individual tasks learning and base model meta-updating [7].

In the inner-loop optimization, for $i^{th}$ task, $\boldsymbol{\theta}_i$ is initialized with $\boldsymbol{\theta}^*$ and adapted over $(\mathbf{X}^t_{i,a}, \mathbf{Y}^t_{i,a})$ with $J$ gradient descent updates, where one gradient descent update with learning rate $\alpha$ is performed by

$$\boldsymbol{\theta}_i \leftarrow \boldsymbol{\theta}_i - \alpha \nabla_{\boldsymbol{\theta}_i} L(\mathbf{Y}^t_{i,a}, f(\mathbf{X}^t_{i,a}; \boldsymbol{\theta}_i)) \qquad (2)$$

In the outer-loop optimization, we employ the first-order MAML [7] to update the meta-model $\boldsymbol{\theta}^*$ with learning rate $\eta$ by using the loss of adapted models over the validation data

$$\boldsymbol{\theta}^* \leftarrow \boldsymbol{\theta}^* - \eta \frac{1}{I} \sum_{i=1}^{I} \nabla_{\boldsymbol{\theta}_i} L(\mathbf{Y}^v_{i,a}, f(\mathbf{X}^v_{i,a}; \boldsymbol{\theta}_i)) \qquad (3)$$

MAML integrates the adaptation process into its framework by first simulating speaker-specific adaptation, then meta-updates the model $\boldsymbol{\theta}^*$. As a result, MAML optimizes for adaptation to dysarthric speakers, which produces a good initialization that can be rapidly adapted to unseen dysarthric speakers.

### 3.3. Reptile based model re-initialization

Reptile differs from MAML with a simpler yet effective meta-model updating manner [8] by removing gradients calculation in the outer-loop optimization. Specifically, for $i^{th}$ task, i.e., adaptation to $i^{th}$ speaker, only training data $(\mathbf{X}^t_{i,a}, \mathbf{Y}^t_{i,a})$ is sampled from the adaptation data, then inner-loop optimization is performed like MAML to obtain $\boldsymbol{\theta}_i$. After obtaining all adapted models $\boldsymbol{\theta}_i$ ($i=1, 2, ..., I$), the meta-model $\boldsymbol{\theta}^*$ is updated via

$$\theta^* \leftarrow \theta^* - \eta \frac{1}{I} \sum_{i=1}^{I} \left( \theta^* - \theta_i \right) \quad (4)$$

$\theta^* - \theta_i$ can be treated as the 'gradient' contributed by $i^{th}$ speaker, averaging gradients of all speakers steers the meta-learning to find an initialization near all speakers' solution manifolds [8]. As a result, the re-initialized model $\theta^*$ can well generalize to unseen dysarthric speakers and perform fast adaptation. Note that when each task is adapted with one step, Reptile is remarkably similar with joint training. However, when multiple adaptation steps are employed, the update incorporates second and higher derivatives that facilitate the generalization [8]. To show the priority of Reptile, the joint training-based model re-initialization is conducted and compared in our experiments.

### 3.4. Per-task batch normalization

In our experiments, the batch normalization [21] is used in the end-to-end ASR model, it aims to re-center and re-scale the intermediate features $h$ of hidden layers via

$$h \leftarrow \gamma \frac{h - \mu}{\sqrt{\sigma^2 + \varepsilon}} + \beta \quad (5)$$

where $\mu$ and $\sigma^2$ are the mean and variance statistics estimated on the current batch or as running statistics, $\gamma$ and $\beta$ are trainable scale and offset weights, $\varepsilon$ is a constant to avoid division by 0. In the original implementation of MAML [7], $\mu$ and $\sigma^2$ are the estimated statistics for the current batch, which results in batch normalization being less effective for speaker adaptation, as the parameters are adapted for statistics that are different from statistics used during the inference. To alleviate this issue, [19] employed batch renormalization [22] with running statistics to renormalize the intermediate features, and running statistics are updated and shared by different speakers during the inner-loop optimization of MAML. However, our preliminary experiments found this way may cause the meta-learning to fail to converge. Since different dysarthric speakers have significantly different acoustic characteristics, the shared running statistics are not suitable to normalize intermediate features of different speakers.

To stabilize the meta-learning, we propose to use per-task batch normalization running statistics, i.e., each adaptation task uses and updates its own running statistics in the inner-loop optimization. In the real implementation, $I$ sets of running statistics are retained for $I$ tasks respectively. As a result, each task learning converges in a stable manner. Besides, the employment of running statistics increase the convergence speed and generalization performance [23]. To be consistent with individual tasks learning during the meta-learning, the re-initialized model uses the batch normalization statistics of the base model for speaker adaptation to unseen dysarthric speakers.

## 4. Experiments

### 4.1. Experimental settings

#### 4.1.1. Dataset

UASpeech [9] and Librispeech [24] datasets are used in our experiments. Librispeech only contains normal speech, and we employ 960h training data for training and dev-clean data as validation data to select best base model $\theta$. UASpeech contains 16 dysarthric speakers and 13 normal speakers, where only the dysarthric speech is used for experiments. Based on subjective speech intelligibility ratings, dysarthric speakers are grouped into 4 severity levels: Severe, Moderate Severe, Moderate and Mild. The dysarthric speech of each speaker has 3 blocks. Following the previous data splits [10], we utilize blocks 1 and 3 as adaptation data for model re-initialization and speaker adaptation, and block 2 as testing data.

#### 4.1.2. ASR Models

Two different end-to-end architectures are used for ASR models: (1) Listen, Attend and Spell (LAS) model [25], which is an attention based sequence-to-sequence model that maps acoustic features to characters. LAS contains an encoder (a 6-layer VGG extractor followed by 5-layer BLSTM with 512 units per direction), 300-dimensional location-aware attention [26] and a decoder that is a single-layer LSTM with 512 units. (2) QuartzNet [27], which is an improved version of Jasper model [28] and achieves comparable performance to Jasper but with much less parameters. The QuartzNet contains 15 blocks with 5 sub-blocks per block, each sub-block consists of a 1-D separable convolution, batch normalization, ReLU and dropout. The inputs of ASR models are 64-dimensional mel-spectrogram, and the targets are 29-dimensional (26 English lower-case letters, a punctuation symbol, a space and a blank token).

#### 4.1.3. Training and evaluation

For the base models training, LAS is trained with the combined CTC and CE loss by using Adadelta optimizer [29] with learning rate of 1 and batch size 16, QuartzNet is trained with CTC loss by using Novagrad optimizer [28] with learning rate of 0.05 and batch size 16. The final base LAS and QuartzNet achieves 10.5% and 6.8% word error rate (WER) on dev-clean split of Librispeech with greedy decoding.

For the model re-initialization, leave-one-subject-out cross validation is adopted, i.e., all dysarthric speakers except for the target speaker are used for MAML/Reptile based model re-initialization, while the speech of the target speaker is used for adaptation and testing. In the inner-loop optimization, updating steps $J$ for both LAS and QuartzNet are 5 with learning rate $\alpha$ of 1e-4. In the outer-loop optimization, the updating steps $K$ for both LAS and QuartzNet are 150, while the learning rate $\eta$ is 1 and 3e-4 for LAS and QuartzNet with the same optimizers as the base models training, respectively. For comparison with Reptile, joint training based model re-initialization with adaptation data of all dysarthric speakers is implemented, where the same learning rate and optimizers as the outer-loop optimization of MAML are used, the model is re-initialized for only 1 epoch, as no improvements are observed with more epochs.

For the speaker adaptation, the total adaptation epochs for each speaker is 10. For LAS, the learning rate of Adadelta remains unchanged as 1. For QuartzNet, the initial learning rate of Novagrad is 3e-4, then decayed by a half every epoch when the adaptation epoch is larger than 5.

For the testing, we adopt weighted finite-state transducers based decoding [30] by incorporating the lexicon and language models into CTC/CE decoding with beam search size as 5, where the lexicon and uniform language models are generated based on the transcriptions of UASpeech following previously published work [10].

To show the effectiveness of the proposed methods, we also demonstrate the results achieved by a hybrid DNN-HMM system [13], which is one of the state-of-the-art dysarthric speech recognition systems on UASpeech dataset.

### 4.2. Experimental results

#### 4.2.1. Performance of different models and strategies

Table 1 shows the performance comparison of different combinations of models and strategies. First, we can observe that compared with base models, the end-to-end ASR benefits from the speaker adaptation, and dramatic improvements can be achieved especially for speakers with Severe/Mod.Severe/Moderate dysarthria, which shows the significance of speaker adaptation for achieving good ASR performance. Second, QuartzNet outperforms LAS under the same strategy condition, we attribute it to that the base model with lower WER has more powerful capacity to generalize to dysarthric speakers and facilitates the adaptation process to obtain better fine-tuned models. Third, the proposed MAML/Reptile based model re-initialization improves the adaptation performance with lower overall WER, where the best system (QuartzNet+Reptile) achieves 54.2% and 7.6% relative WER reduction compared with the base model without finetuning and direct adaptation on the base model, respectively, which shows that meta-learning based model re-initialization can effectively acquire speech knowledge across dysarthric speakers, which enhances the generalization capacity of the re-initialized model to unseen dysarthric speakers. Besides, joint training based model re-initialization can only improve the adaptation performance for speakers with Severe or Mod.Severe dysarthria, as the joint training aims to find a global minimizer of CTC/CE loss of all dysarthric speakers, and hence focuses more on minimizing the larger losses of speakers with more severe dysarthria, leading to better initializations for those speakers. Finally, we can observe that most end-to-end ASR systems can achieve comparable performance with the hybrid DNN-HMM system, which can be even outperformed by the base QuartzNet with Reptile, this further demonstrates the effectiveness of the proposed MAML/Reptile based model re-initialization to achieve good dysarthric speech recognition performance.

Table 1. *WER (%) of different combinations of models and strategies, strategies are followed by the base models: '+Adapt' means perform speaker adaptation based on base models, '+M & Adapt' means first perform M-based model re-initialization and then speaker adaptation based on the re-initialized model*

| Systems | Severity levels | | | | Overall |
|---|---|---|---|---|---|
| | Severe | Mod.Severe | Moderate | Mild | |
| DNN-HMM [13] | 67.4 | 31.4 | 23.6 | 12.2 | 30.6 |
| Base LAS | 98.8 | 90.7 | 82.9 | 47.2 | 76.4 |
| +Adapt | 73.7 | 43.2 | 34.9 | 13.0 | 37.9 |
| +Joint & Adapt | **67.8** | 41.9 | 36.2 | 19.2 | 38.6 |
| +MAML & Adapt | 70.0 | **38.2** | **31.1** | 13.2 | 35.2 |
| +Reptile & Adapt | 68.7 | 39.0 | 32.5 | **12.2** | **35.0** |
| Base QuartzNet | 98.0 | 86.4 | 71.3 | 28.8 | 66.6 |
| +Adapt | 76.4 | 36.1 | 24.3 | **7.1** | 33.0 |
| +Joint & Adapt | **65.1** | 35.7 | 32.5 | 14.6 | 36.3 |
| +MAML & Adapt | 70.0 | **33.3** | 24.5 | 7.2 | 30.6 |
| +Reptile & Adapt | 69.3 | 33.7 | **24.1** | 7.3 | **30.5** |

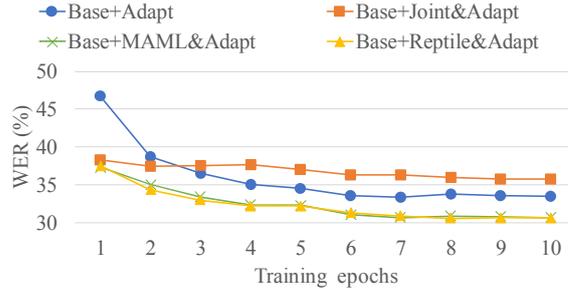

Figure 3. *WER (%) versus training epochs for QuartzNet*

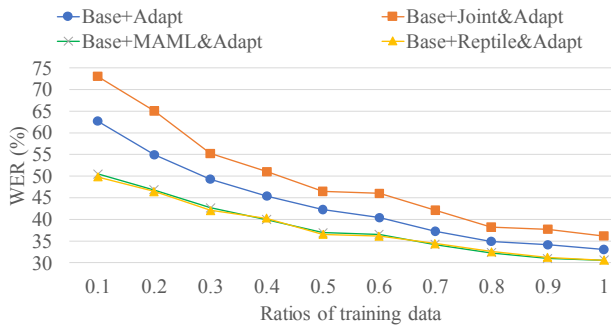

Figure 2. *WER (%) versus ratios of training data for QuartzNet*

#### 4.2.2. Impact of training data size

To verify the advantages of the proposed methods for speaker adaptation with limited data, we vary the amount of training data, i.e., different ratios of total adaptation data are used for finetuning. WER results based on QuartzNet are reported in Figure 2, we can observe that the WER of all methods decreases with the increased training data as expected, the proposed MAML and Reptile based methods achieve the similar and lower WER in all cases. Compared with 'Base+Adapt' and 'Base+Joint&Adapt', significant WER reduction can be achieved by the proposed methods when the ratio is small with fewer training data, this shows the MAML/Reptile based re-initialization enables the model to effectively acquire cross-dysarthric-speaker knowledge, which facilitates the speaker adaptation to improve the ASR performance with limited data.

#### 4.2.3. Learning curves

In addition, we demonstrate the learning curves of different methods in Figure 3, i.e., the WER versus training epochs. We can observe that the performance of 'Base+Joint&Adapt' nearly saturates in the first several training epochs, as the data for joint training based model re-initialization is also used for finetuning, fast convergence can be achieved for speaker adaptation. Contrarily, without using the data of target dysarthric speaker for model re-initialization, MAML and Reptile based methods can achieve fast adaptation and lower WER compared with 'Base+Adapt' and 'Base+Joint&Adapt', which validates the effectiveness of the proposed methods for obtaining an appropriate model initialization with powerful generalization capacity to unseen dysarthric speech.

### 5. Conclusions

This paper focuses on applying meta-learning to find a good end-to-end model initialization to dysarthric speech recognition. Given a base model pre-trained from large scale normal speech data, MAML and Reptile are proposed to meta-update the base model by incorporating across-dysarthric-speakers knowledge into the re-initialized model, which allows rapid speaker adaptation to unseen dysarthric speakers. Speaker adaptation results on UASpeech dataset show that compared with the base LAS and QuartzNet models, the best re-initialized models can achieve 7.7% and 7.6% relative WER reduction, respectively. Our future work focuses on extending the proposed methods for hybrid system, e.g., DNN-HMM, to achieve better performance.

### 6. Acknowledgements

This work is partially supported by the General Research Fund from the Research Grants Council of Hong Kong SAR Government (Project No. 14208817).